\documentstyle[12pt]{article}

\newfont{\set}{msbm10 scaled \magstep 1}
\newfont{\Set}{msbm6}

\font\tenmsa=msam10 scaled 1700
\def\hook{\hbox{\tenmsa\char'171}\ }

\newcommand{\R}{$\mbox{\set R}$}

\newcommand{\be}{\begin{equation}}
\newcommand{\ee}{\end{equation}}
\newcommand{\bea}{\begin{eqnarray}}
\newcommand{\eea}{\end{eqnarray}}
\newcommand{\tr}{{\rm tr}}
\newcommand{\lag}{{\cal L}}

\newcommand{\var}{{\cal E}}

\newcommand{\ldv}{{\sf L}}

\newcommand{\vq}{{\bf v}_{Q}}

\newcommand{\vf}{{\bf v}}
\newcommand{\pr}{{\rm pr}\,}
\newcommand{\jet}{{\rm J}}
\newcommand{\prf}{{\rm pr}}
\newcommand{\Df}{{\sf D}}

\def\pd{\partial}

\makeatletter

\def\eqnarray{\stepcounter{equation}\let\@currentlabel=\theequation
\global\@eqnswtrue
\global\@eqcnt\z@\tabskip\@centering\let\\=\@eqncr
$$\halign to \displaywidth\bgroup\@eqnsel\hskip\@centering
  $\displaystyle\tabskip\z@{##}$&\global\@eqcnt\@ne
  \hfil$\displaystyle{\hbox{}##\hbox{}}$\hfil
  &\global\@eqcnt\tw@ $\displaystyle\tabskip\z@
  {##}$\hfil\tabskip\@centering&\llap{##}\tabskip\z@\cr}
\@addtoreset{equation}{section}
\def\theequation{\thesection.\arabic{equation}}
\makeatother

\begin{document}

\begin{titlepage}

\date{19th September 1995}
\title{Properties of the Scalar Universal Equations}

\author{J.A. Mulvey \\
        {\it Department of Mathematical Sciences} \\
        {\it University of Durham,} \\
        {\it South Road,} \\
        {\it Durham, DH1 3LE, UK} \\
        email: {\tt J.A.Mulvey@durham.ac.uk}}

\maketitle

{\centerline {\bf Preprint DTP/95/53}}

\begin{abstract}
The variational properties of the scalar so--called ``Universal''
equations are reviewed and generalised. In particular, we note that
contrary to earlier claims, each member of the Euler hierarchy may
have an explicit field dependence. The Euler hierarchy itself is given
a new interpretation in terms of the formal complex of variational
calculus, and is shown to be related to the algebra of distinguished
symmetries of the first source form.
\end{abstract}

\end{titlepage}

\section{Introduction \label{ufeintro}}

The Universal Field models proposed in \cite{univ1,univ2,univ3,univ4}
are a class of integrable field theories with a wide variety of
attractive features:
\begin{itemize}
\item they may be formulated in an arbitrary number of space--time
  dimensions;
\item they are either diffeomorphism or reparametrisation invariant in
  the dependent variables;
\item they are Lorentz and Euclidean invariant (and in fact $GL(m,\R)$
  invariant) in the independent variables;
\item they are derived from an infinite number of inequivalent Lagrangians.
\end{itemize}
Furthermore, the scalar versions of the theory have been shown to be
linearisable. One of the scalar theories is a direct generalisation of
the Bateman equation \cite{meme} and is linearisable by a Legendre
transformation in the manner described in \cite{univ4}.

The business at hand is to describe and explain the strange
variational properties of the scalar Universal equations. Consider a
scalar field $\phi$ dependent on $m$ space(--time) coordinates.  The
variational derivation of the Universal equations presented in
\cite{univ3} is based on the idea of the {\it generic Euler
hierarchy}.  Consider a Lagrangian density \(\lag_0\) which only
depends on the first derivatives of $\phi$. Compute the variation of
this and construct a new function,
\be
\lag_1 = F_1 \var \lag_0,
\label{l1}
\ee
where $\var$ is the Euler variation and $F_1$ is some real--valued
function dependent only on the first derivatives of $\phi$. Then
consider $\lag_1$ to be the Lagrangian for a new equation, $\var
\lag_1$. Continue the process to arrive at:
\be
\lag_k = F_k \var \lag_{k-1}.
\label{lk}
\ee
This sequence terminates at the $(m+1)$th stage: $\var \lag_{m} =0$.
Furthermore, at the $m$th stage the resulting Euler--Lagrange
expression factorises, and one of the factors is ``universal'' ---
independent of the details of the initial and intermediate
Lagrangians. On setting this Euler--Lagrange form to zero, we arrive
at an equation of motion equivalent to the Monge--Amp\`{e}re
expression,
\be
\Delta_{M-A} = \det H =0,
\label{monge-gen-hier}
\ee
where $H$ is the Hessian matrix of the dependent variable $\phi$: \(
H_{ij} = \phi_{ij} \).

The most interesting special cases of this construction are referred
to in \cite{univ3} as {\it generalised Bateman equations}. The idea is
to use an initial Lagrangian $\lag_0$ which is homogeneous of degree
one in the first derivatives of $\phi$, and to restrict all the $F_k$
to have this property too. Performing the procedure described above
leads to the cessation of the hierarchy a stage earlier than described
above: \(\var \lag_{m-1} = 0 \).  Remarkably, the penultimate
Euler--Lagrange expression $\var \lag_{m-2}$ is again a product, and
again one of the its factors is ``universal''. Setting this expression
to zero gives the Universal Bateman Equation of \cite{univ1} and
sequels.

The equation can be written,
\be
\Delta = \det \left ( \begin{array}{clcl} 0 & \phi_{x_1} & \ldots &
\phi_{x_m} \\ \phi_{x_1} & \phi_{x_1 x_1} & \ldots & \phi_{x_1 x_m} \\
\vdots & \vdots & \ddots & \vdots \\ \phi_{x_m} & \phi_{x_m x_1} &
\ldots & \phi_{x_m x_m} \end{array} \right )=0,
\label{u-detform}
\ee
or in components,
\be
\Delta = \varepsilon_{i_1 \ldots i_m}\varepsilon_{j_1 \ldots j_m}
\phi_{i_1} \phi_{j_1} \phi_{i_2 j_2} \ldots \phi_{i_m j_m}=0.
\label{u-compform}
\ee
The component form makes it easy to verify that $\Delta$ can also be
expressed as:
\be
\Delta =  \tr \left (GH^{\dagger}\right ),
\label{neat}
\ee
where the matrix $G$ has components
\be G_{ij}= \phi_{i}\phi_{j}
\ee
and
\be
H^{\dagger} ={\rm adj} ( H )
\label{aij}
\ee
is the classical adjoint matrix of $H$. Alternatively, we could define
a new matrix $U$ such that (\ref{neat}) is equivalent to the equation:
\be
\Delta = \tr  (U H).
\label{neat2}
\ee
The explicit form of $U$ is easily deduced from
(\ref{u-compform}). It is:
\be
U= \varepsilon_{i_1 \ldots i_m}\varepsilon_{j_1 \ldots j_m}
\phi_{i_1} \phi_{j_1} \phi_{i_2 j_2} \ldots \phi_{i_{m-1} j_{m-1}}.
\label{def-u-alpha}
\ee
A supplementary assumption is that $\det H$ is non-vanishing.

Our mission is to try to understand and generalise the generic and
Bateman hierarchies described above from the point of view of the
standard theory of variational calculus as presented in, for example,
the book by Olver \cite{olver}. In particular, we wish to know under
what circumstances we can introduce dependence on the field $\phi$
into the initial Lagrangian and the multiplier functions $F_k$.  We
will find that only in certain circumstances does the Euler hierarchy
define a ``universal'' theory.  In addition, we will think about what
the iterative procedure means in terms of the formal variational
complex, described in \cite{olver} or the work of Anderson
\cite{anderson1}. We will find that the iterated equations of motion
are related to determining equations for distinguished symmetries of
higher members of the hierarchy, and we will justify this connection
with a calculation of the generalised symmetries of the Universal
equations.

Basic references (apart from the original papers
\cite{univ1,univ2,univ3,univ4}) are Olver \cite{olver}, the review
articles by Anderson \cite{anderson1,anderson2} and his recent work
with Pohjanpelto \cite{anderson3,anderson4}. Another view of the
universal equations and their variational symmetry properties is given
by Grigore \cite{grig1,grig2}.

\section{Lagrangian Properties \label{u-lagr-prop}}
We begin by analysing the Euler hierarchy in the language of
variational calculus presented in \cite{olver,anderson1}. Initially,
we will only assume that the initial Lagrangian depends on the field
and its first derivatives. In the language of jet bundles
\cite{Saunders}, we ask that the Lagrangian is a smooth function
\(\lag_0:\jet^1 \pi \rightarrow \R\), where \(\jet^1 \pi\) is the first
jet bundle of the bundle $\pi$.  In this case we are restricting our
attention to the trivial bundle $\pi:\R^{m+1} \rightarrow \R^m$. We
make no assumptions about any symmetries of the resulting action. This
language will be useful later.

{}From $\lag_0$, we derive an Euler--Lagrange form:
\be
\Delta_1 = \Df_{\lag_0 }^{*} (1) = \var \lag_0.
\label{delta1}
\ee
{}From this we construct the next Lagrangian,
\be
\lag_1=F_1 \Delta_1,
\label{lagr1}
\ee
where we assume that $F_1$ depends only on $\phi$ and its first
derivatives. The next source form follows readily using the product
formula:
\be
\var ({\cal F}_1 {\cal F}_2 ) = \Df_{{\cal F}_1}^{*} ({\cal F}_2) +
\Df_{{\cal F}_2}^{*} ({\cal F}_1),
\label{fairytale}
\ee
where $\Df^{*}$ indicates the adjoint of the Fr\'{e}chet derivative
\cite{olver}. In this case:
\be
\Delta_2 = \var \lag_1 =
\Df_{\Delta_1}^{*} (F_1) + \Df_{F_1}^{*} (\Delta_1).
\label{ald1}
\ee
Now, the Helmholtz condition of the calculus of variations states that
an equation is an exact Euler--Lagrange form if and only if its Fr\'{e}chet
derivative is self--adjoint. Hence:
\bea
\Df_{\Delta_1}^{*} (F_1)
& = & \Df_{\Delta_1} (F_1) \nonumber \\ & = & \frac{\pd \Delta_1}{\pd
\phi_J}D_J (F_1)
\label{fstterm1}
\eea
The second term on the right--hand side of (\ref{ald1}) just
turns out to be,
\be
\Df_{F_1}^{*} (\Delta_1) = \Delta_1 \frac{\pd
F_1}{\pd \phi} - D_i \left ( \frac{\pd F_1}{\pd \phi_i} \Delta_1
\right ),
\label{sndterm1}
\ee
since $F_1$ only depends on $\phi$ and its first derivatives.

Putting (\ref{fstterm1}) and (\ref{sndterm1}) into (\ref{ald1}), a
short calculation tells us that:
\bea
\Delta_2 & = & \Delta_1 \var F_1 + \frac{\pd \Delta_1}{\pd \phi}
\left ( F_1 - \frac{\pd F_1}{\pd \phi_i} \phi_i \right ) + \frac{\pd
\Delta_1}{\pd \phi_i} \frac{\pd F_1}{\pd \phi} \phi_i \nonumber \\
& & + \frac{\pd \Delta_1}{\pd \phi_{ij}} \left ( D_i \left ( \frac{\pd
F_1}{\pd \phi} \phi_j \right ) + D_i \left ( \frac{\pd F_1}{\pd
\phi_k} \right ) \phi_{jk} \right ).
\label{delta2}
\eea
The important point to notice is that this construction guarantees
that there are no derivatives of order higher than two in the
resulting equation of motion. This means that when the process is
continued, the analysis is essentially unchanged. We are led to the
iterative formula for the $(k+1)$th Euler--Lagrange form:
\bea
\Delta_{k+1} & = & \Delta_{k} \var F_{k} + \frac{\pd \Delta_{k}}{\pd
\phi} \left ( F_{k} - \frac{\pd F_{k}}{\pd \phi_i} \phi_i \right ) +
\frac{\pd \Delta_{k}}{\pd \phi_i} \frac{\pd F_{k}}{\pd \phi} \phi_i
\nonumber \\
&  & + \frac{\pd \Delta_{k}}{\pd \phi_{ij}} \left ( D_i
\left ( \frac{\pd F_{k}}{\pd \phi} \phi_j \right ) + D_i \left (
\frac{\pd F_{k}}{\pd \phi_l} \right ) \phi_{jl} \right ).
\label{iter-form}
\eea
There is no such recursive definition if $F_k$ depends on second or higher
derivatives.

We are interested in sequences generated by this kind of recursion
which terminate after a finite number of iterations. The expression
(\ref{iter-form}) simplifies greatly if we restrict attention to $F_k$
that are {\it (1)} independent of $\phi$, and {\it (2)} homogeneous of
degree one in the first derivatives. Then we find:
\be
\Delta_{k+1}  =  \Delta_{k} \var F_{k} + \frac{\pd \Delta_{k}}{\pd
\phi_{ij}} D_i \left ( \frac{\pd F_{k}}{\pd \phi_l} \right ) \phi_{jl},
\label{ourcase}
\ee
or more symmetrically,
\be
\Delta_{k+1} = \left ( \frac{\pd
\Delta_{k}}{\pd \phi_{kl}} \phi_{ik} \phi_{jl} - \Delta_k \phi_{ij}
\right ) \frac{\pd^2 F_k}{\pd \phi_i \pd \phi_j}.
\label{ourcase-rewrite}
\ee

This is precisely the recurrence found by Fairlie and Govaerts
(\cite{univ3}) in their treatment of the generic hierarchy. They found
that if $\lag_0$ and all the $F_k$ were independent of $\phi$ then the
sequence terminated in a Monge--Amp\`{e}re equation. This sequence
only produces such a result if it is assumed that the initial
Lagrangian is independent of $\phi$, in which case the condition {\it
(2)} can be relaxed anyway, so there is no new information. Otherwise,
(\ref{ourcase}) fails to reproduce the generic hierarchy. Termination
of the hierarchy seems to depend on certain linear-algebraic
properties induced by the fact that the second derivatives appear
homogeneously at each stage of the sequence \cite{univ3}. This does
not happen if the $F_k$ depend on $\phi$. So far, all attempts to
generalise the construction of the generic hierarchy by relaxing this
requirement have failed. Sample computer calculations carried out on
MAPLE show that such constructions do not terminate in universal,
exact source forms at either the $(m-1)$th or $m$th stages.

We proceed to generate the equations of the generic hierarchy under
the restriction that the $\lag_0$ and $F_k$ are $\phi$-independent.
We know that the first equation (\ref{delta1}) has the explicit form,
\be
\Delta_1=-\frac{\pd^2 \lag_0}{\pd \phi_i \pd \phi_j}\phi_{ij},
\label{bog-standard}
\ee which can be written:
\be
\Delta_1=-\tr \left ( H M_0 \right ),
\label{sl-less-cr}
\ee
where $M_0$ is the Hessian matrix of $\lag_0$ with respect to its
dependence on first derivatives of $\phi$. It is then a
straightforward matter to apply the recursion relation
(\ref{ourcase-rewrite}) to this starting term. Using the notation,
\bea
\left [ M_k \right ]_{ij} & = & \frac{\pd^2 F_k}{\pd \phi_i \pd
\phi_j}, \nonumber \\ P_k & = & H M_k,
\label{not-loc}
\eea
the first few terms of the hierarchy are:
\bea
\Delta_1 & = &
-\tr \left (P_0 \right ), \nonumber \\ \Delta_2 & = & \tr \left (P_0
\right )\tr \left (P_1 \right ) - tr \left (P_0 P_1 \right ),
\nonumber \\ \Delta_3 & = & -\tr \left (P_2 \left ( P_0 P_1 + P_1 P_0
\right ) \right ) + \tr \left (P_0 \right ) \tr \left (P_1 P_2 \right
) + \tr \left (P_1 \right ) \tr \left (P_0 P_2 \right ) \nonumber \\ &
+ & \tr \left (P_2 \right ) \tr \left (P_0 P_1 \right ) - \tr \left
(P_0 \right ) \tr \left (P_1 \right ) \tr \left (P_2 \right ),
\nonumber \\ \Delta_4 & = & -\tr \left( P_3 \left( P_1 P_2 P_0 + P_0
P_1 P_2 + P_2 P_0 P_1 + P_0 P_2 P_1 + P_1 P_0 P_2 + P_2 P_1 P_0 \right
) \right ) \nonumber \\ & + & \tr \left ( P_1 P_2 \right ) \tr \left (
P_0 P_3 \right ) + \tr \left ( P_2 P_3 \right ) \tr \left ( P_0 P_1
\right ) + \tr \left ( P_1 P_3 \right ) \tr \left ( P_0 P_2 \right )
\nonumber \\ & + & \tr \left ( P_0 \right ) \left ( \tr \left ( P_1
P_2 P_3 \right ) + \tr \left ( P_3 P_2 P_1 \right ) \right ) + \tr
\left ( P_1 \right ) \left ( \tr \left ( P_0 P_2 P_3 \right ) + \tr
\left ( P_3 P_2 P_0 \right ) \right ) \nonumber \\ & + & \tr \left (
P_2 \right ) \left ( \tr \left ( P_0 P_1 P_3 \right ) + \tr \left (
P_3 P_1 P_0 \right ) \right ) \nonumber \\ & - & \tr \left ( P_0
\right ) \tr \left ( P_1 \right ) \tr \left ( P_2 P_3 \right ) - \tr
\left ( P_0 \right ) \tr \left ( P_2 \right ) \tr \left ( P_1 P_3
\right ) - \tr \left ( P_1 \right ) \tr \left ( P_2 \right ) \tr \left
( P_0 P_3 \right ) \nonumber \\ & - & \tr \left ( P_3 \right )
\Delta_3 \nonumber \\ &\vdots &
\label{expl-tow}
\eea
The subsequent members of the hierarchy become combinatorially
more complicated.

We can use the recursion relation as the basis for a plausibility
argument that demonstrates the vanishing of the $\Delta_k$ at a
certain stage, and hence when the $F_k$ is the characteristic of a
conservation law.  We wish to know when $\Delta_k$ vanishes for
{\it any} possible choices of $F_k$.  (Of course, we are only interested in
those cases when the Hessian of $F_k$ is non-vanishing.) This amounts
to solving the matrix differential equation,
\be
H\Gamma^k H = \Delta_k H,
\label{exact-con-mat}
\ee where, \be \Gamma^{k}_{ij} = \frac{\pd \Delta_k}{\pd \phi_{ij}}.
\label{gamma-def}
\ee The equation (\ref{exact-con-mat}) is solved by any $\Delta_k$ of
the form $\Delta = ({\rm some\ factor})\times \det H$, where ``some
factor'' is independent of the second derivatives of $\phi$. We know from
the variational calculation that the second derivatives only enter the
source forms such that $\Delta_k$ is homogeneous of degree $k$ in the
second derivatives. Therefore, this solution can only work at the
$m$th stage. We will see in the next section that any function of the
first derivatives of $\phi$ is the characteristic of a generalised
symmetry of $\Delta_{M-A}$. It follows immediately that taking an
Euler variation of $\Delta_{M-A}$ as described above will give a zero result.

A more rigorous proof of these results is presented in \cite{univ3}
using an explicit calculation.

This completes the discussion of the generic case. Now we turn to the
Bateman case. The reader should keep in mind Euler's theorem on
homogeneous functions, which states that, if ${\cal F}({\bf x})$ is
homogeneous of degree $\alpha$ in its arguments ${\bf x} = (x_1, ...,
x_n)$, then:
\be
\sum_{i=1}^{n}
x_{i} \frac{\pd {\cal F}}{\pd x_i}=\alpha {\cal F}.
\label{eulerth}
\ee
In our case, we assume that $\lag_0$ is homogeneous of degree one in
the first derivatives of $\phi$, meaning that,
\be
\phi_i \frac{\pd \lag_0}{\pd \phi_i} = \lag_0.
\label{h-lag}
\ee
It is easy to derive further homogeneity properties of the Euler
variation and the derivatives of $\lag_0$.

In particular, the first derivative of $\lag_0$ with respect to
$\phi_i$ is homogeneous of degree zero:
\be
\phi_i \frac{\pd^2 \lag_0}{\pd \phi_i \pd \phi_j} = 0,
\label{sing-eq}
\ee
which implies the singularity condition:
\be
\det (M_0) = 0,
\label{sing-lag}
\ee
since $\phi$ is arbitrary.

So we proceed to discussion of the Bateman hierarchy. Now we will
assume that the initial Lagrangian and the $F_k$ have arbitrary
dependence on $\phi$ and are homogeneous of degree one in first
derivatives of $\phi$. This greatly civilises the shapes of both
$\Delta_1$ and $\Delta_2$ from their original forms (\ref{delta1}) and
(\ref{delta2}). If we use the various properties stemming from Euler's
theorem, we recalculate $\Delta_1$ and $\Delta_2$ to be,
\bea
\Delta_1 & = & -\frac{\pd^2 \lag_0}{\pd \phi_i
\phi_j}\phi_{ij}, \nonumber \\ \Delta_2 & = & \left ( \frac{\pd
\Delta_{1}}{\pd \phi_{kl}} \phi_{ik} \phi_{jl} - \Delta_1 \phi_{ij}
\right ) \frac{\pd^2 F_1}{\pd \phi_i \pd \phi_j},
\label{hom-magic}
\eea
from which we deduce precisely the same recurrence
(\ref{ourcase-rewrite}) without the restriction that the $F_k$ need be
independent of $\phi$.

The recursive procedure now defines a set of equations identical in
form to (\ref{expl-tow}), and so the proof of Fairlie and Govaerts
\cite{univ4} is still valid. The equation governing the termination of
the sequence (\ref{exact-con-mat}) is satisfied by (\ref{u-compform}).

There is a nice interpretation of all this in terms of the theory of
the Euler--Lagrange complex studied in \cite{anderson1}. Let
$\Omega^{*}$ denote the algebra of differential forms on the infinite
jet bundle $\jet^{\infty}\pi$. This algebra is bigraded into,
\be
\Omega^{*} =  \bigoplus_{r,s} \Omega^{(r,s)},
\label{bigrad}
\ee
where the spaces $\Omega^{(r,s)}$ contain forms with the local
coordinate expression:
\[  f[\phi]\ dx^{i_1}\wedge \ldots \wedge dx^{i_r}\wedge
\theta^{\alpha_{1}}_{J_1} \wedge \ldots \wedge
\theta^{\alpha_{s}}_{J_s}, \]
the forms $\theta$ being the various contact one--forms. The various
differentials and subcomplexes of this space are discussed in
\cite{anderson1}. Alternatively, see the book by Saunders
\cite{Saunders}.

The starting Lagrangian is a form $\lambda_0\in \Omega^{(m,0)}$. The
corresponding variation is the source form $\Delta_1 d\phi \wedge
\omega =\var (\lambda)\in \Omega^{m,1}$. Now consider taking a Lie
derivative (denoted by $\ldv$) of $\Delta_1$ with respect to the
prolongation of some vector field $X_1$ whose evolutionary form is
$X_{F_1}$, with characteristic $F_1$. For an arbitrary equation
\(\Delta \), we have the Cartan formula:
\be
\ldv_{\pr X_{1}} (\Delta \ d\phi \wedge \omega) = \delta_V \left (
X_{F_1} \hook \Delta \ d\phi \wedge \omega
\right ) + X_{F_1} \hook \delta_V \left ( \Delta \ d\phi \wedge \omega
\right ).
\label{var-comp-lie}
\ee
The differentials $\delta_V$ are those of the {\it Euler--Lagrange
complex} defined in \cite{olver} or \cite{anderson1}. A proof of
(\ref{var-comp-lie})  can be found in \cite{anderson3}.  Each term has
a simple interpretation in the calculus of
variations. The left hand side vanishes identically if $X_1$ is a
distinguished symmetry of $\Delta$. The first term on the right is an
Euler variation, and so it vanishes if $F_1$ is the characteristic
of a conservation law. Finally, the $\delta_V$ in the third term is
just a Helmholtz operator, so it vanishes if $\Delta$ is an exact
Euler--Lagrange form. Not surprisingly, this is the central formula in
the recent studies of the generalised Noether theorem by Anderson and
Pohjanpelto \cite{anderson3,anderson4}.

Returning to our example, (\ref{var-comp-lie}) takes on a particularly
simple form when applied to $\Delta_1$. Since $\Delta_1$ is an exact
Euler--Lagrange form, the Helmholtz term vanishes and we are left with
the form $\Delta_2$ as defined in (\ref{ald1}). This interpretation
holds for all $\Delta_k$, so we can view the Euler hierarchy as
repeated application of the Lie derivative to successive source forms.
So the recursive definition becomes:
\be
\Delta_{k+1} \ d\phi \wedge \omega = \ldv_{\prf \vf_{F_k}} \left (
\Delta_k \ d\phi \wedge \omega \right ).
\label{rec-lie-der}
\ee
The ``universal'' theory rests on the observation that at a certain
stage the all source forms so defined, in either the generic or
Bateman hierarchy, are equivalent to one another and their flows
defined by the Lie derivative are identically zero.  In such a
situation, the iteration vanishes identically, and this yields a
product expansion of the type (\ref{fairytale}) which is equal to zero.
This indicates that $\lag_k$ is a total divergence, and hence $F_k$
must be the characteristic of a conservation law. Explicitly, we
deduce that any function of the first derivatives of $\phi$ is a
generalised symmetry of the Monge--Amp\`{e}re equation and that any
function of homogeneous of weight one in the $\phi_i$ but with
arbitrary dependence on $\phi$ is a generalised symmetry of the
Bateman--type Universal equation (\ref{u-detform}). These results will
be proved explicitly in the next section.

Changing emphasis, we can view the $F_k$ as unknown functions to be
determined, and then we can interpret $\Delta_2$ as the equation
determining the distinguished symmetry algebra of the Euler--Lagrange
form $\Delta_1$. The third and subsequent source forms
$\Delta_3, \ldots, \Delta_m$ are a set of nested equations determining
distinguished symmetry algebras for their immediate predecessors. It
would be interesting to know what information, if any, can be gleaned
from these higher equations about the symmetries of the original
equation $\Delta_1$.

\section{Generalised Symmetries \label{gsmd-sec}}
In order to confirm our understanding of how the generic and Bateman
hierarchies terminate, it will be helpful to know more about the
generalised symmetries of the equations (\ref{monge-gen-hier}) and
(\ref{u-detform}). To that end we will carry out a detailed analysis
of each of each equation in the manner described in Olver's book
\cite{olver}. We will look for first--order generalised symmetries ---
in other words symmetries whose evolutionary characteristic depends on
the $x_i$, $\phi$ and the first derivatives of $\phi$. The reader is
reminded of the standard expansions of the total derivatives $D_i$ and
$D_{ij}$:
\bea
D_i F & = & \frac{\pd F}{\pd x_i} +\frac{\pd F}{\pd \phi_J} \phi_{Ji},
\nonumber \\ D_{ij} F & = & \frac{\pd^2 F}{\pd x_i \pd x_j} +
\frac{\pd^2 F}{\pd x_i \pd \phi_K}\phi_{Kj} + \frac{\pd F}{\pd
\phi_J}\phi_{Jij} + \frac{\pd^2 F}{\pd x_j \pd \phi_J} \phi_{Ji}
+\frac{\pd^2F}{\pd \phi_J \pd \phi_K }\phi_{Ji}\phi_{Kj},
\eea
where the multi-indices $J,K$ will actually only have length zero or
one in the cases we are considering.

We will begin with the equation (\ref{monge-gen-hier}). It turns out
that the first--order generalised symmetries of this equation span a
rather large infinity of possibilities. Using the standard procedure,
we seek a generalised symmetry in the evolutionary form,
\be
\vq = Q\frac{\pd}{\pd \phi}.
\label{evolform}
\ee
The symmetry equation is quite straightforward:
\be
\prf \vq ( \Delta_{M-A} ) = \frac{\pd
\Delta_{M-A}}{\pd \phi_{ij}} D_{ij} Q[\phi] = H^{\dagger}_{ij} D_{ij}
Q[\phi].
\label{m-a-sym-con}
\ee
The symbol $\prf$ denotes the infinite prolongation. On expanding the
total derivative, we find that the coefficients of the third
derivatives of $\phi$ vanish on solutions of (\ref{monge-gen-hier})
(after taking into account the first prolongation of the equation of
motion) and the first symmetry constraint comes from the term of order
$(m+1)$ in the second derivatives, which gives:
\be
H^{\dagger}_{ij} \frac{\pd^2 Q}{\pd
\phi_k \pd \phi_l} \phi_{ik} \phi_{jl} = 0,
\label{m-a-qu-2-der}
\ee
on solutions. Writing,
\be \left [ H_Q \right ]_{kl} = \frac{\pd^2
Q}{\pd \phi_k \pd \phi_l},
\label{hq-def}
\ee
we find that (\ref{m-a-qu-2-der}) becomes:
\be \tr \left (
H^{\dagger} H H_Q H \right ) = \det H \ \tr \left ( H_Q H \right ) =
0,
\label{m-a-cru}
\ee
and this is identically true on solutions of $\Delta_{M-A}=0$.
This pattern continues for the terms of order $m$ in second
derivatives, and we find that a $Q$ with arbitrary dependence on the
first derivatives is a generalised symmetry for
(\ref{monge-gen-hier}). This is the expected result given the
interpretation of the generic hierarchy and confirms its termination
at the $(m+1)$th stage.

Now we turn our attention to the symmetry algebra of the Universal
equation (\ref{u-detform}), which turns out to be a little more
restrictive than in the generic case. The determining equation is,
\be
\frac{\pd \Delta}{\pd \phi_i}(D_i Q[\phi])+\frac{\pd \Delta}{\pd
\phi_{(ij)}} (D_{ij} Q[\phi]) =0,
\label{univ-g-s-eq}
\ee
on solutions of $\Delta = 0$. Again, we will only look for
first--order generalised symmetries. This should be sufficient to
confirm our interpretation of the Bateman hierarchy in the previous
section. Furthermore, experience with the two dimensional case
\cite{meme} suggests that the symmetries dependent on first derivatives
are responsible for the linearisability property.

Given this assumption, (\ref{univ-g-s-eq}) expands to become,
\be
\varepsilon_{i_1 \ldots i_m}\varepsilon_{j_1 \ldots j_m}
\phi_{i_1}\phi_{j_1}\phi_{i_2 j_2} \ldots \phi_{i_m j_m} +
(m-1) U_{kl} \delta_{(ij)(kl)}D_{ij} Q  = 0,
\label{exp-g-s-eq}
\ee
where,
\be U_{kl} = \varepsilon_{i_1 \ldots
i_{m-1}k}\varepsilon_{j_1 \ldots j_{m-1}l} \phi_{i_1}\phi_{j_1}\phi_{i_2 j_2}
\ldots \phi_{i_{m-1} j_{m-1}}
\label{uhdef}
\ee
is the sort of matrix appearing in (\ref{neat2}).

Following the usual algorithm, we must set the left side of
(\ref{exp-g-s-eq}) to zero order--by--order in the derivatives, taking
into account the equation $\Delta = 0$. Using the first prolongation
of $\Delta$, we find that third order derivatives automatically
vanish, and so the first task is to find the condition for the
vanishing of terms of order $m$ in the second derivatives of $\phi$.

Extracting the relevant term from (\ref{exp-g-s-eq}), we define a
matrix $S$ such that:
\bea S_{kl}\frac{\pd^2 Q}{\pd \phi_k \pd \phi_l}
& = & (m-1) H_{ki}U_{i j} H_{j l} \frac{\pd^2 Q}{\pd \phi_k \pd
\phi_l} \nonumber \\
\ & = & 0,
\label{qu-con}
\eea
on solutions of $\Delta =0$. To simplify this, we use the form
(\ref{neat2}) to rewrite $\Delta$ using the cyclicity and linearity of
trace and the properties of the classical adjoint:
\be
\Delta = \frac{1}{m} \frac{1}{\det H} \tr \left (
HU H H^{\dagger} \right ),
\ee
and then, by associativity,
\be
\Delta = \frac{1}{m(m-1)}\frac{1}{\det H} \tr \left (
SH^{\dagger} \right ).
\ee
On comparison with (\ref{neat}) we find
that $S=m(m-1)\det H\ G$ and the ``on--shell'' symmetry condition is
satisfied by any $S=$(some scalar factor)$\times \det H\ G$. This
produces a symmetry condition,
\be \phi_i \phi_j \frac{\pd^2 Q}{\pd
\phi_i \pd \phi_j} =0,
\label{gen-fogs-con}
\ee
which is satisfied by any $Q$ homogeneous of degree zero or one in the
first derivatives of $\phi$. This confirms the facts deduced earlier
from the Bateman hierarchy.

According to a theorem of Kumei and Bluman \cite{kum-blu}, the
existence of a generalised symmetry which depends on the solution to a
linear equation such as (\ref{gen-fogs-con}) guarantees the existence
of a linearising transform for (\ref{u-detform}). This is indeed the
case, as demonstrated in \cite{univ4}.

Having disposed of the term with $m$ second derivatives, we need to
equate the term with $m-1$ second derivatives to zero. The relevant
equation can be written as:
\be
H^{\dagger}_{ij}\frac{\pd Q}{\pd x_i}\phi_j + 2 (m-1)
U_{kl} \delta_{(ij)(kl)} \left ( \frac{\pd^2
Q}{\pd x_i \pd \phi_n} \phi_{jn} + \frac{\pd^2 Q}{\pd \phi \pd \phi_n}
\phi_i \phi_{jn} \right ) = 0.
\label{g-c-n-1}
\ee
This can be solved by a characteristic of the form,
\be
Q=g(\phi)F(\eta),
\label{char-of}
\ee
where $g$ is an arbitrary smooth, real--valued function, $\eta =
x_i \phi_i$ (no sum) and $F$ is a smooth, real--valued function which
respects the homogeneity properties that we have decided on for $Q$.
Notice that the diffeomorphism symmetry of the equation is included in
this solution.

Finally, to get rid of the remaining terms, we use (\ref{char-of}) and
arrive eventually at the condition:
\be
\varepsilon_{i_1 \ldots i_m}\varepsilon_{j_i
\ldots j_m} \phi_{i_1} \phi_{j_1} \phi_{i_2 j_2} \ldots
\phi_{i}\phi_{j} \ldots \phi_{i_m j_m} \left (
g F^{\prime \prime} + 2 g^{\prime} F^{\prime} +g^{\prime \prime}F
\right )=0.
\label{n-2-con}
\ee This is identically true due to the antisymmetry of the
$\varepsilon$ symbol.

\section{Conclusion}
The generic Euler hierarchy and the special case of the Bateman
hierarchy have been interpreted as a sequence of iterated Lie
derivatives for the distinguished symmetries of a large class of
Lagrangians. It has been shown that the Bateman hierarchy alone admits
explicit dependency on the field due to its homogeneity properties,
but the form of the generic hierarchy is unchanged from the analysis
in \cite{univ4}. The termination of both hierarchies is guaranteed by
the generalised symmetries of their associated universal equations,
which in turn imply the existence of an infinite class of conservation
laws for each equation.

Many questions remain unanswered about the geometrical meaning of the
iteration of the Lie derivatives, and the presence of any helpful
algebraic properties. It also remains to apply the procedure to more
general geometrical constructions than the trivial bundle $\pi$
considered here. It is noted that other equations follow from similar
procedures, namely certain multidimensional generalisations of the
Born--Infeld equation \cite{mulv}.

\vspace{1cm}

\noindent {\large {\bf Acknowledgements}} \newline
\noindent I am very grateful to D.B. Fairlie for much advice and
encouragement. This work was financially supported by the Department of
Education for Northern Ireland.

\end{document}